\documentclass[preprint,showpacs,aps]{revtex4}

\usepackage{epsfig}

\begin{document}

\title{Critical domain-wall dynamics of model B}

\author{R.H. Dong, B. Zheng \footnote{corresponding author; email: zheng@zimp.zju.edu.cn} and N.J. Zhou}

\affiliation{$^1$ Zhejiang University, Zhejiang Institute of Modern
                  Physics, Hangzhou 310027, P.R. China}

\begin{abstract}
With Monte Carlo methods, we simulate the critical domain-wall
dynamics of model B, taking the two-dimensional Ising model as an
example. In the macroscopic short-time regime, a dynamic scaling
form is revealed. Due to the existence of the quasi-random walkers,
the magnetization shows intrinsic dependence on the lattice size
$L$. A new exponent which governs the $L$-dependence of the
magnetization is measured to be $\sigma=0.243(8)$.
\end{abstract}

\pacs{64.60.Ht, 68.35.Rh, 05.10.Ln}

\maketitle

According to Hohenberg and Halperin \cite{hoh77}, a dynamic system
with no conservation laws is called model A, while that with a
conserved order parameter is model B. For {\it critical} systems,
the equilibrium state of model B is in a same universality class of
model A. For lattice models, such as the Ising model, the critical
temperature of model B is also the same as that of model A. However,
the dynamic universality class of model B is different from that of
model A. For the two-dimensional (2D) $\phi^4$ theory, one has
derived that the dynamic exponent $z=4-\eta$, with $\eta$ being the
well known static exponent \cite{hoh77}. In other words, $z$ is {\it
not} an independent one, and its value is around $4$, much larger
than $z \approx 2$ of model A.

The dynamic scaling behavior of model A has been understood around
and even {\it far from} equilibrium
\cite{hoh77,jan89,hus89,zhe98,zhe99}. Based on the dynamic scaling
form in the {\it macroscopic} short-time regime, new methods for the
determination of both dynamic and static critical exponents have
been developed \cite{li95,luo98,zhe98}. Recent progress in the
short-time critical dynamics includes, for example, theoretical
calculations and numerical simulations of the XY models and
Josephson junction arrays \cite{zhe03a,oze03b,gra05,liu08}, magnets
with quenched disorder \cite{yin05,oze05,che05}, ageing phenomena
\cite{cal05,lei07,rom08}, and various applications and developments
\cite{lee05,ara07,fan07,lin08,bag08}. Very recently, critical
relaxation of a domain wall has been concerned \cite{zho08,he09},
and it is also relevant for the domain-wall dynamics at zero or low
temperatures \cite{nat01,kle07a}.

The dynamics of model B is important in various fields. For example,
the phase ordering dynamics of model B describes the spinodal
decomposition of binary alloys and phase separation of fluids
\cite{bra94,bra02}. The dynamics of driven lattice gases also
belongs to model B \cite{alb02,car05,bag08,smi08}. However, the {\it
critical} dynamics of model B is less studied in the literature,
compared to that of model A \cite{cal05}. This is mainly because its
critical slowing down is more severe, and it is difficult to reach
the equilibrium. From this view, it is instructive to explore the
short-time critical dynamics of model B, for the critical slowing
down does not disturb so much the simulations
\cite{li95,luo98,zhe98}. In Ref. \cite{alb02}, the short-time
dynamic scaling form is applied to numerically identify the
universality classes of anisotropic driven lattice gases. Since the
dynamic system is rather complicated, it induces a controversy
\cite{car04,alb04}.

Our thought is to clarify the short-time dynamic scaling form of
model B starting from simpler systems. For the 2D Ising model,
critical relaxation with a {\it disordered} initial state has been
simulated with the Kawasaki algorithm
\cite{ale94,maj94,god04,sir04}. The results support that the dynamic
exponent is $z=4-\eta=15/4$. To fully understand the short-time
critical dynamics of model B, we should explore the dynamic effects
of different initial conditions. An ordered state can not be the
initial state of model B. But the dynamic relaxation with a {\it
semi-ordered} initial state is important. In fact, it describes the
dynamic evolution of a domain wall. In the case of model A, there
emerge plenty of new phenomena \cite{zho08,he09,nat01,kle07a,zho09}.

The purpose of this paper is to investigate the critical domain-wall
dynamics of model B with Monte Carlo methods. To be specific, we
simulate the dynamic relaxation of the 2D Ising model starting from
a semi-ordered state with the Kawasaki algorithm. The semi-ordered
initial state consists of two fully-ordered domains with opposite
spin orientations. As time evolves, the domain wall between two
domains roughens, and looks like a growing interface, i.e., the
so-called {\it domain interface}. Such a dynamic process is
inhomogenous in space, very different from that with a disordered
initial state.

In Monte Carlo simulations, the 2D Ising model is defined on a
rectangular lattice, with a linear size $2L_x$ in the $x$ direction
and $L_y$ in the $y$ direction. {\it Anti-periodic} and periodic
boundary conditions are adopted in the $x$ and $y$ directions
respectively. Initially, spins are taken to be positive on the
sublattice $L_x\times L_y$ at the right side and negative at the
left side. With the semi-ordered initial state, we update the spin
configuration by exchanging two spins in the nearest neighbor with
the heat-bath algorithm at the critical temperature $T_c$, up to a
maximum time $t_M = 10^6$. Most simulations are performed with $L_x=
L_y=L$. The total of samples for average is $30\ 000$, $20\ 000$,
$3000$ and $2000$ for $L = 32$, $64$, $128$ and $256$ respectively.
Statistical errors are estimated by dividing the samples into two or
three subgroups. If the fluctuation in the time direction is
comparable with or larger than the statistical error, it will be
taken into account. For convenience, we set the $x$-axis such that
the domain wall between the positive and negative spins is located
at $x=0$. So the $x$ coordinate of a lattice site is a half-integer.

For simplicity, we first set $L_x= L_y=L$. Due to the semi-ordered
initial state, the time evolution of the dynamic system is
inhomogeneous in the $x$ direction. Therefore we measure the
magnetization as a function of $x$ and $t$,
\begin{equation}
M(t,x,L) = \frac{1}{L} \left \langle \left [\sum^L_{y =
1}S_{xy}(t)\right ] \right \rangle.  \label{equ10}
\end{equation}
Here $S_{xy}(t)$ is the spin at the time $t$ on the lattice site
$(x,y)$, and $<\ldots>$ represents the statistical average.
Generally, the magnetization may depend on $L$.

At the critical temperature, there are three spatial scales in the
dynamic system, i.e., the non-equilibrium correlation length
$\xi(t)$, the coordinate $x$ and the lattice size $L$. One may
believe that $\xi(t)$ grows in a universal form $\xi (t) \sim
t^{1/z}$ in all spatial directions, because of the homogeneity of
the interactions in the Hamiltonian. Therefore, standard scaling
arguments lead to the scaling form of the magnetization
\begin{equation}
M(t,x,L)=t^{-\beta /\nu z} \widetilde{M}(t^{1/z}/x,t^{1/z}/L),
\label{equ20}
\end{equation}
where $\beta$ and $\nu$ are the static exponents, and $z$ is the
dynamic exponent. For the 2D Ising model with the dynamics of model
B, theoretical values of the exponents are $\beta=1/8$,  $\nu=1$ and
$z=15/4$. On the right side of the above equation, the overall
factors $t^{-\beta / \nu z}$ indicates the scaling dimension of $M$,
and the scaling function $\widetilde{M}(s,u)$ describes the scale
invariance of the dynamic system. In the macroscopic long-time
regime, i.e., the regime with $\xi(t)\ge L$, it has been well known
that the dynamic system exhibits a universal dynamic scaling
behavior. In this paper, we assume that the scaling form in
Eq.~(\ref{equ20}) holds already in the {\it macroscopic} short-time
regime, i.e., the regime with $\xi(t)\ll L$, after a microscopic
time scale $t_{mic}$.

\begin{figure}[h]
\epsfysize=5.5cm \epsfclipoff \fboxsep=0pt
\setlength{\unitlength}{1.cm}
\begin{picture}(10,6)(0,0)
\put(-3.2,-0.1){{\epsffile{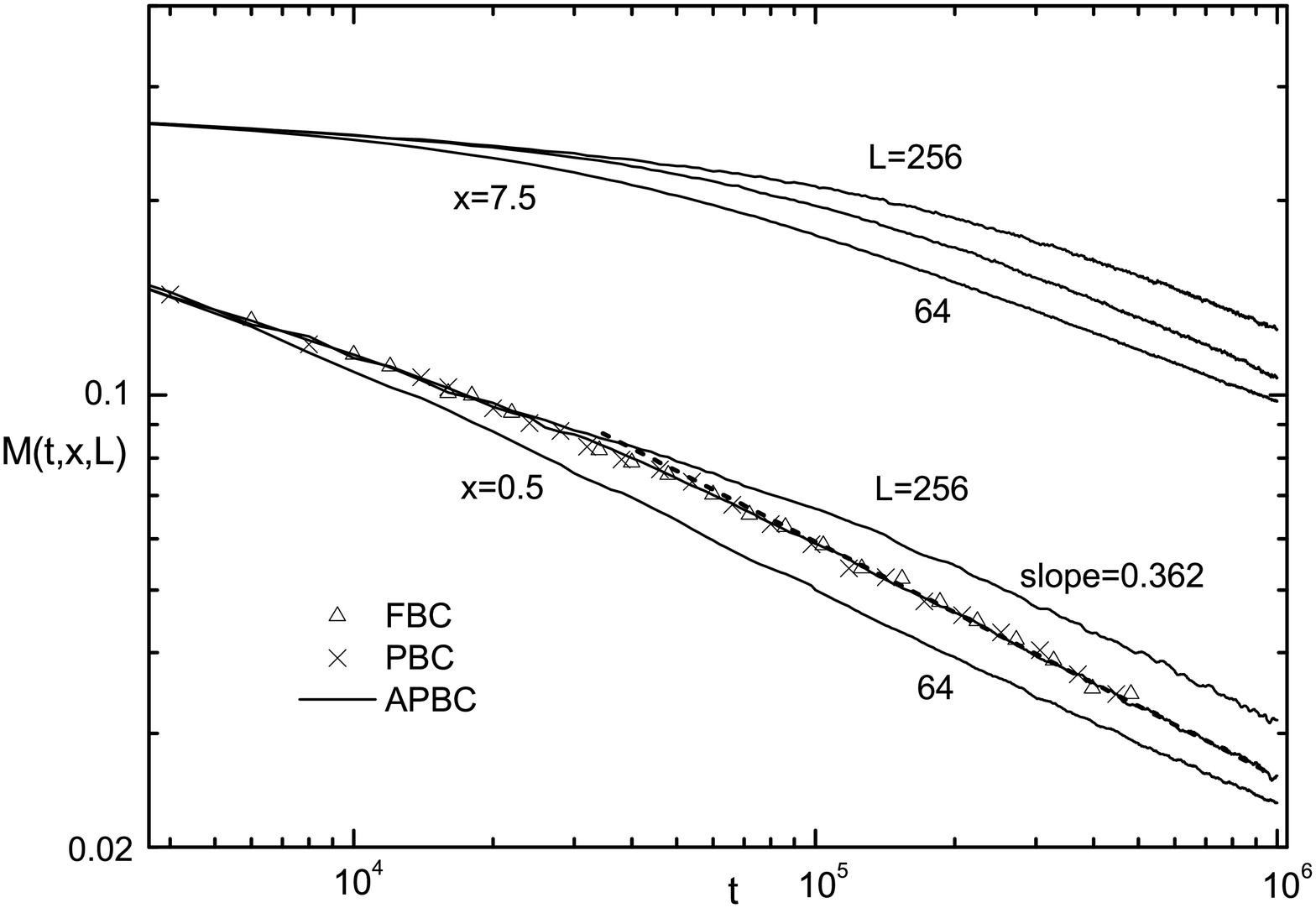}}}\epsfysize=5.5cm
\put(5.3,-0.1){{\epsffile{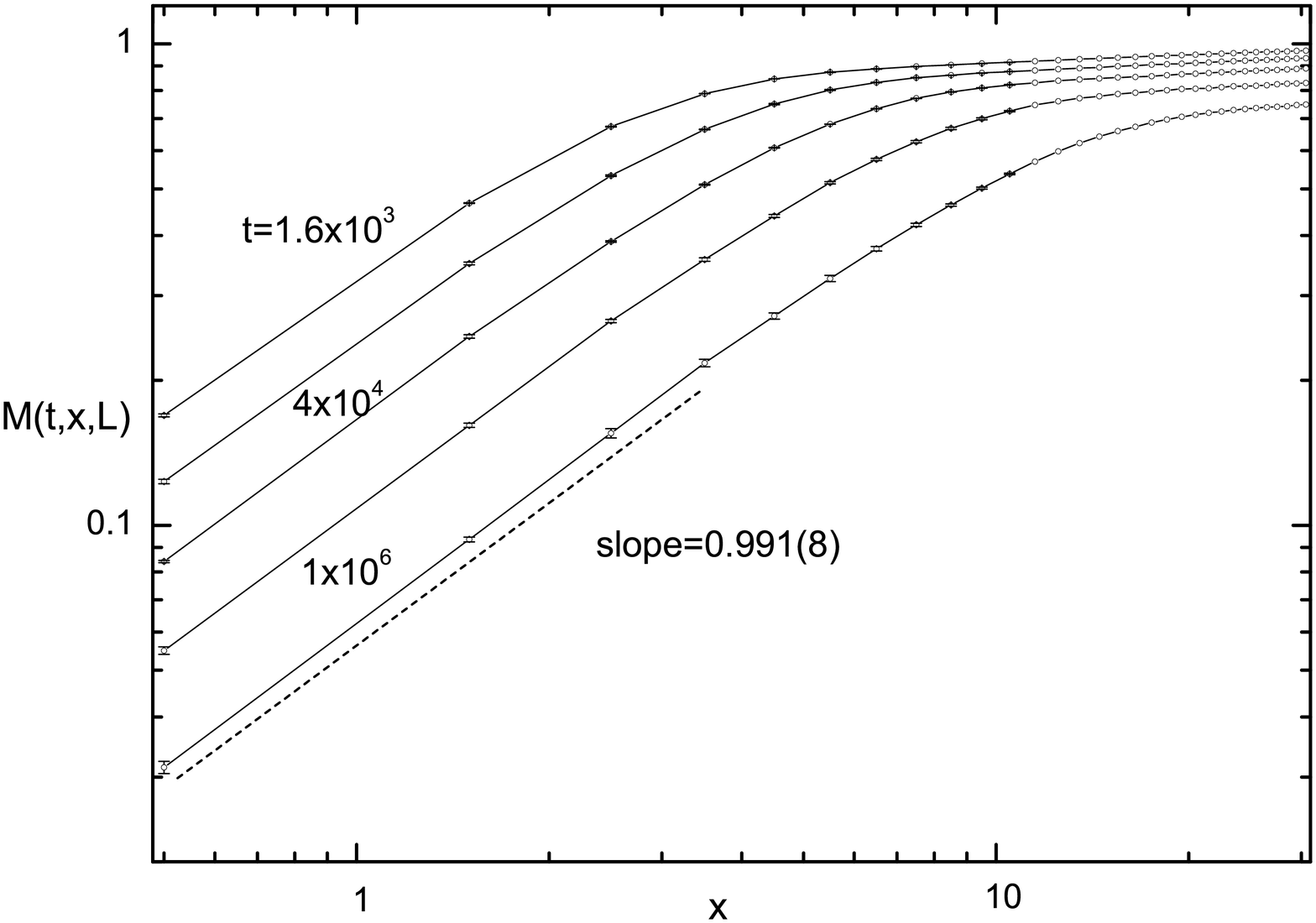}}}
\put(-2.4,-0.3){\footnotesize{(a)}}\put(6.2,-0.3){\footnotesize{(b)}}
\end{picture}
\caption{(a) The time evolution of the magnetization is plotted for
different $x$, and $L=64$, $128$ and $256$ (from below) on a
double-log scale. An anti-periodic boundary condition (APBC) is
adopted in the $x$ direction. For comparison, results of $x=0.5$ and
$L=128$ with periodic and free boundary conditions (PBC and FBC) are
also included. The dashed line shows a power-law fit. (b) The
magnetization obtained with L=256 is plotted versus $x$ for
different $t$ on a double-log scale. Errors of the data points are
less or about one percent. The dashed line shows a power-law fit.}
\label{fig1}
\end{figure}

For the dynamics of model A, the dynamic scaling behavior of the
magnetization is relatively simple. In the short-time regime, the
finite-size effect of $M(t,x,L)$ is negligible, because of
$\xi(t)\ll L$. $M(t,x,L) \to t^{-\beta /\nu z} F(s)$ with
$s=t^{1/z}/x$. For a large $s$, i.e., inside the domain interface,
$F(s)$ exhibits a power-law behavior $F(s) \sim s^{-\beta_0/\nu}$
with $\beta_0/\nu \approx 1$. For a small $s$, i.e., outside the
domain interface, $F(s) \sim const $ \cite{zho08}. In the long-time
regime, the magnetization may depend on the lattice size $L$,
typically dominated by an exponential law $M(t,L) \sim exp(-t/L^z)$.
For the dynamics of model B, the dynamic scaling behavior of the
magnetization is rather complicated. It is anomal that {\it the
magnetization shows intrinsic dependence on the lattice size $L$
even in the short-time regime}. In Fig.~\ref{fig1} (a), the time
evolution of the magnetization is displayed. For a fixed $x$,
$M(t,x,L)$ obviously varies with $L$. Although $M(t,x,L)$ strongly
depends on $L$, it is insensitive to the boundary condition. In
Fig.~\ref{fig1} (a), numerical simulations with periodic and free
boundary conditions in the $x$ direction are also included. All
boundary conditions lead to the same results.

The domain-wall motion of model B is driven by interchanging
positive and negative spins in neighbor. When a negative spin
escapes from the domain interface and jumps into the positive
domain, it becomes {\it quasi-free} and moves {\it randomly}. We
call this spin '{\it a quasi-random walker}'. The average moving
distance of the quasi-random walkers in a time $t$ is the order of
$l_r(t) \sim \sqrt{t}$. Even in the short-time regime, i.e., the
regime with $\xi(t) \sim t^{1/z}\ll L$, it may occur $L<l_r(t)\sim
\sqrt{t}$, for the dynamic exponent $z=15/4 > 2$. Therefore the
quasi-random walkers easily touch the boundary of the dynamic
system. This should be the physical origin of the $L$-dependence of
the magnetization in the short-time regime.

In fact, the average moving distance $l_r(t)$ of the quasi-random
walkers is an additional spatial scale for the dynamics of model B.
In the literatures, such a diffusion length scale $l_r(t) \sim
t^{\phi/z}$ with $\phi \approx z/2$ has been detected in the dynamic
relaxation with a disordered initial state \cite{sir04,cal05}. In
principle, the magnetization should also depend on the scaling
variable $l_r(t)/L$. The dynamic scaling form in Eq.~(\ref{equ20})
holds only when the quasi-random walkers reach a 'homogeneous'
state, i.e., $l_r(t)\gg L$. In other words, the quasi-random walkers
should touch the boundary, turn back to the domain interface, and
then stabilize at a homogenous state. At $x=0.5$, for example, the
shortest time $t_S$ for this movement is about $t_S =4L^2$ in our
numerical simulations.

Since $M(t,x,L)$ relies on $t$ through two scaling variables in
Eq.~(\ref{equ20}), its dynamic behavior is complicated.
Nevertheless, a dynamic scaling form is typically characterized by
power-law behaviors. Let us now concentrate on the following
features of the scaling function $\widetilde{M}(s,u)$,
\begin{equation}
   \widetilde{M}(s,u) \to \{
   \begin{array}{lll}
    G(u) s^{-\beta_0 / \nu}  & \quad &  \mbox{ $s\to \infty$} \\
  F(s)u^{-\sigma}   & \quad & \mbox{ $s\to \infty$,} \ \mbox{$u\to 0$}
   \end{array}
\label{equ30}
\end{equation}
with $s=t^{1/z}/x$ and $u=t^{1/z}/L$. Actually, $F(s)=s^{-\beta_0 /
\nu}$. We use the notation $F(s)$ for convenience. The exponent
$\beta_0/\nu$ characterizes the spatial behavior of the
magnetization inside the domain interface, which appears similar to
the surface exponent defined on a surface \cite{zho08,lin08,he09}.
Thus we call it the interface exponent. $\sigma$ is a new exponent
exclusive for the dynamics of model B, for describing the
finite-size dependence of the magnetization.

In Fig. \ref{fig1} (b), the magnetization is plotted versus $x$.
Inside the domain interface, i.e., for $x\ll \xi(t)$ or a large $s$,
$M(t,x,L)$ exhibits a power-law behavior. The slope of the curves is
$0.991(8)$, very close to $1$. This suggests $\beta_0/\nu=1$, and
indicates that $M(t,x,L)$ is an analytic function of $x$, the same
as that of model A \cite{zho08}. In Fig. \ref{fig1} (a), we detect a
power-law decay of the magnetization inside the domain interface in
the short-time regime, i.e., for large $s$ and small $u$. The slope
is estimated to be $0.362$ from the curve of $L=128$. For $L=64$, it
is reaching the long-time regime for $t > 10^5$, and a deviation
from the power law is observed. This power-law behavior indicates
$(\beta/\nu +\beta_0/\nu+\sigma)/z \approx 0.362$, and yields
$\sigma \approx 0.242$ with  the input $\beta/\nu=1/8$ and $z=15/4$.
Since $\sigma$ is {\it positive}, the magnetization in Eq.~(\ref
{equ30}) increases with $L$. Therefore we call this $L$-dependence
of the magnetization {\it intrinsic}.

\begin{figure}[h]
\epsfysize=5.4cm \epsfclipoff \fboxsep=0pt
\setlength{\unitlength}{1.cm}
\begin{picture}(10,6)(0,0)
\put(-3.3,-0.1){{\epsffile{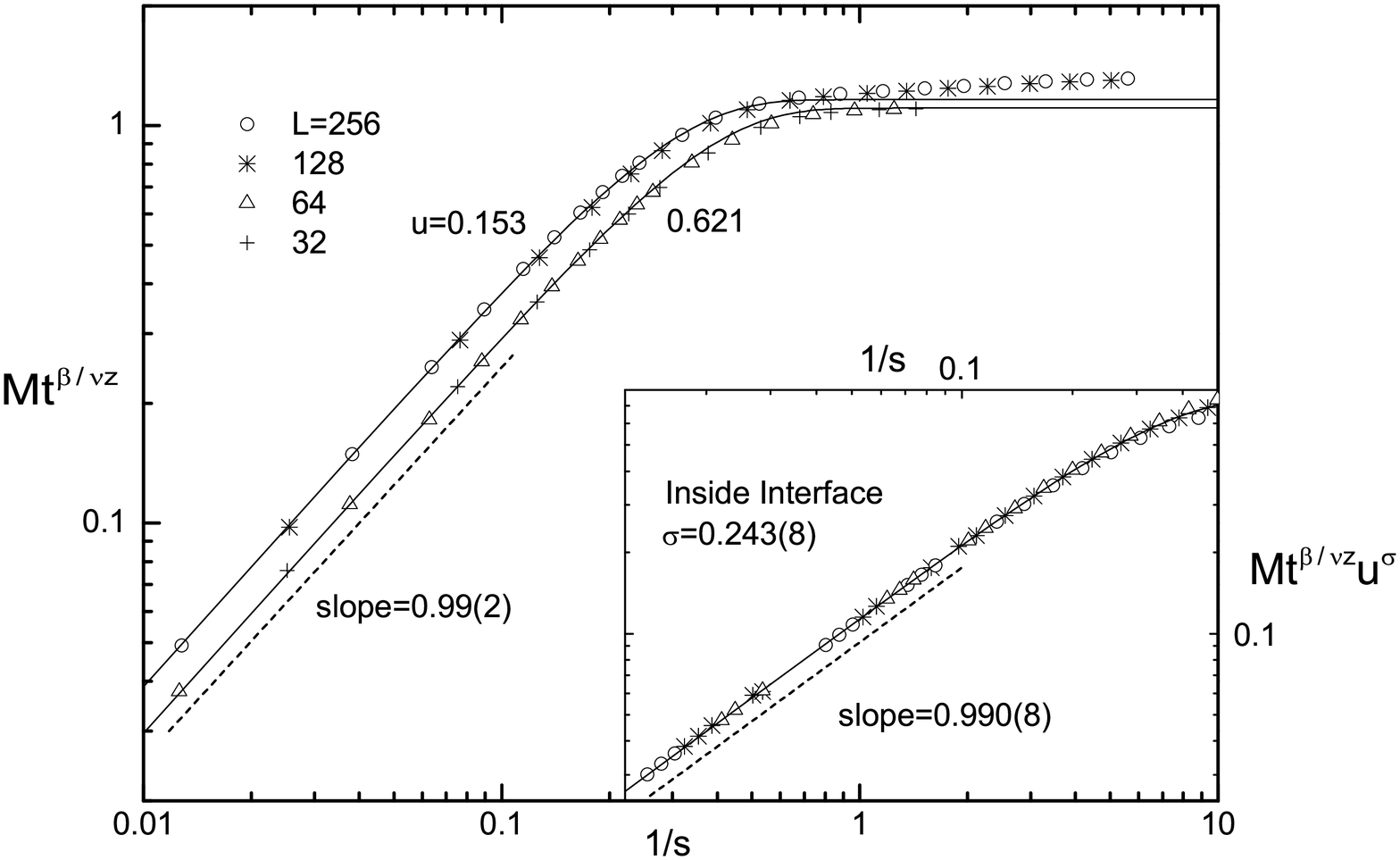}}}\epsfysize=5.5cm
\put(5.,-0.1){{\epsffile{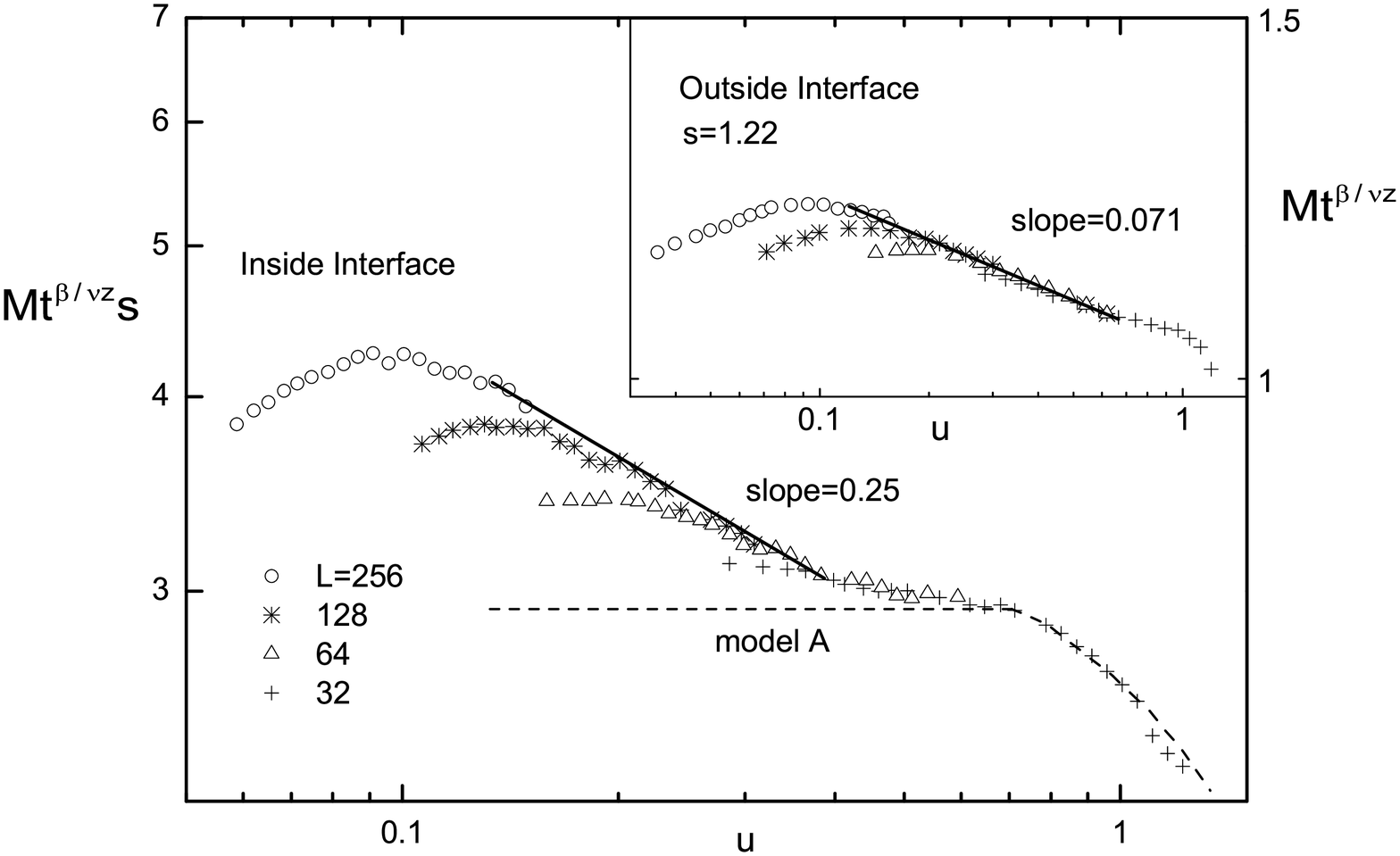}}}
\put(-2.5,-0.3){\footnotesize{(a)}}\put(6.,-0.3){\footnotesize{(b)}}
\end{picture}
\caption{(a) The scaling function $M(t,x,L)t^{\beta /\nu
z}=\widetilde{M}(s,u)$ with $s=t^{1/z}/x$ and $u=t^{1/z}/L$ is
plotted versus $1/s$ for fixed $u=0.53$ and $0.621$. Data collapse
is observed for different $L$. Solid lines represent the error
function $f(y) \sim \int^y exp(-x^2)dx$, and the dashed line shows a
power-law fit.  In the inset, a power-law behavior
$\widetilde{M}(s,u) \sim F(s) u^{-\sigma}$ is assumed in the
short-time regime, and data of different $u$ and $L$ inside the
domain interface collapse onto a single curve $Mt^{\beta /\nu
z}u^{\sigma} \sim F(s)$. (b) A power-law behavior
$\widetilde{M}(s,u) \sim G(u) s^{-1}$ is assumed inside the domain
interface, and data of different $L$ collapse onto the master curve
$Mt^{\beta /\nu z}s \sim G(u)$. The solid line indicates the
power-law behavior in the short-time regime. Departure of the data
from the master curve in the time regime $t < t_S\sim 4 L^2$ is also
displayed. The dashed line schematically shows the scaling function
of model A. In the inset, the scaling function $\widetilde{M}(s,u)$
outside the domain interface is plotted versus $u$.} \label{fig2}
\end{figure}

To illustrate the scaling form comprehensively and precisely, we now
perform scaling plots with Eq.~(\ref{equ20}). In Fig. \ref{fig2}
(a), the scaling function $\widetilde{M}(s,u)$ with a {\it fixed}
$u$ is displayed, and data collapse is observed for different $L$.
In the large-$s$ regime, i.e., inside the domain interface,
$\widetilde{M}(s,u)$ exhibits a power-law behavior $ s^{-\beta_0 /
\nu}$ with $\beta_0 / \nu=0.99(2)$. This agrees with the measurement
in Fig. \ref{fig1} (b). For the dynamics of model A, the scaling
function $\widetilde{M}(s)$ can be fitted to the error function
\cite{zho08,he09}. For the dynamics of model B, $\widetilde{M}(s,u)$
with a fixed $u$ also coincides with the error function in the
large-$s$ regime. However, it does not approach a constant in the
small-$s$ regime, due to the existence of the quasi-random walkers.

According to Eq.~(\ref{equ30}), $\widetilde{M}(s,u) \sim
F(s)u^{-\sigma}$ for large $s$ and small $u$. In the inset of Fig.
\ref{fig2} (a), the scaling plot for $M(t,x,u)t^{\beta/\nu z}
u^{\sigma} \sim F(s)$ is performed. Data collapse is observed inside
the domain interface. With the input $\beta/\nu=1/8$ and $z=15/4$,
we extract the exponents $\sigma=0.243(8)$ and $\beta_0 /
\nu=0.990(8)$, consistent with $0.242$ and $0.991(8)$ estimated from
Fig. \ref{fig1} (a) and (b).

To explicitly reveal the $L$-dependence of the magnetization in
different time regimes, we fix $s$ and plot the scaling function
$\widetilde{M}(s,u)$ versus $u$. Inside the domain interface, i.e,
for a large $s$, $\widetilde{M}(s,u) \sim G(u)s^{-1}$. In Fig.
\ref{fig2} (b), the scaling function $M(t,x,u)t^{\beta/\nu z} s \sim
G(u)$ is displayed at $x=0.5$. In the time regime $t> t_S \sim
4L^2$, data collapse is observed. For a small $u$, there emerges a
power-law decay $G(u) \sim u^{-\sigma}$, and $\sigma$ is estimated
to $0.25$, in agreement with $0.243(8)$ extracted from Fig.
\ref{fig2} (a). For a large $u$, $G(u)$ is governed by an
exponential law. For comparison, a scaling function $G(u)$ of model
A is schematically shown in Fig. \ref{fig2} (b). In the time regime
$t< t_S$, the scaling form is violated due to the quasi-random
walkers, and data of different $L$ do not collapse. To fully
understand the dynamic behavior in this time regime, we need to
include the additional scaling variable $l_r(t)/L$ in the dynamic
scaling form. A detailed description of this kind will be presented
elsewhere.

In the inset of Fig. \ref{fig2} (b), the scaling function of model B
is plotted for a small $s$, i.e., outside the domain interface.
Qualitatively, $G(u)$ is similar to that inside the domain
interface, but the effective $\sigma$ decreases with $s$.

Finally, we should point out that the finite-size dependence of the
magnetization in the domain-wall motion of model B is anisotropic in
spatial directions. In fact, the intrinsic $L$-dependence of the
scaling function $\widetilde{M}(s,u)$ in Eq.~(\ref{equ30}) does
refer only to $L_x$, not $L_y$. Since the domain wall is oriented in
the $y$ direction, the random walkers does not induce anomal
dependence of the magnetization on $L_y$. In the short-time regime,
i.e., the regime with $\xi(t) \ll L_y$, the magnetization is {\it
independent of} $L_y$. In Monte Carlo simulations, we have confirmed
this by fixing $L_x$ and changing $L_y$ from $64$ to $128$ and
$256$.

In summary, we have simulated the critical domain-wall dynamics of
model B with Monte Carlo methods, taking the 2D Ising model as an
example. A short-time dynamic scaling form is revealed, and the
scaling function is carefully computed. Due to the existence of the
quasi-random walkers, the magnetization intrinsically depends on the
lattice size even in the short-time regime. This is very different
from the case of model A. The new exponent $\sigma$ which governs
the $L$-dependence of the magnetization is measured to be
$0.243(8)$. The interface exponent $\beta_0/\nu$ takes the value
$0.99(1)$, close to $1$, the same as that of model A. It is a
challenge to derive the scaling form in Eq.~(\ref{equ20}) and
extract the exponent $\sigma$ in Eq.~(\ref{equ30}) with
renormalization group methods.

{\bf Acknowledgements:} This work was supported in part by NNSF
(China) under grant No. 10875102 and 10325520.


\begin{thebibliography}{10}

\bibitem{hoh77}
{P.C. Hohenberg and B.I. Halperin}, Rev. Mod. Phys. {\bf {49}},  435
(1977).

\bibitem{jan89}
{H.K. Janssen, B. Schaub, and B. Schmittmann}, Z. Phys. {\bf {B
73}},  539  (1989).

\bibitem{hus89}
D. Huse, Phys. Rev. {\bf {B 40}},  304  (1989).

\bibitem{zhe98}
B. Zheng, Int. J. Mod. Phys. {\bf B12},  1419  (1998), review
article.

\bibitem{zhe99}
{B. Zheng, M. Schulz, and S. Trimper}, Phys. Rev. Lett. {\bf {82}},
1891  (1999).

\bibitem{li95}
{Z.B. Li, L. {Sch\"ulke}, and B. Zheng}, Phys. Rev. Lett. {\bf
{74}},  3396  (1995).

\bibitem{luo98}
{H.J. Luo, L. Sch\"ulke, and B. Zheng}, Phys. Rev. Lett. {\bf {81}},
180  (1998).

\bibitem{zhe03a}
{B. Zheng, F. Ren, and H. Ren}, Phys. Rev. {\bf {E68}},  046120
(2003).

\bibitem{oze03b}
{Y. Ozeki and N. Ito}, Phys. Rev. {\bf {B68}},  054414  (2003).

\bibitem{gra05}
{E. Granato and D. Dominguez}, Phys. Rev. {\bf B71},  094521
(2005).

\bibitem{liu08}
{H. Liu, J.P. Lv, and Q.H. Chen}, Europhys. Lett. {\bf 84},  66004
(2008).

\bibitem{yin05}
{J.Q. Yin, B. Zheng, and S. Trimper}, Phys. Rev. {\bf E72},  036122
(2005).

\bibitem{oze05}
{Y. Ozeki and K. Ogawa}, Phys. Rev. {\bf B71},  220407  (2005).

\bibitem{che05}
{Y. Chen and Z.B. Li}, Phys. Rev. {\bf B71},  174433  (2005).

\bibitem{cal05}
{P. Calabrese and A. Gambassi}, J. Phys. {\bf A38},  R133  (2005).

\bibitem{lei07}
{X.W. Lei and B. Zheng}, Phys. Rev. {\bf E75},  040104  (2007).

\bibitem{rom08}
{F. Roma and D. Dominguez}, Phys. Rev. {\bf B78},  184431  (2008).

\bibitem{lee05}
{H.K. Lee and Y. Okabe}, Phys. Rev. {\bf E71},  015102  (2005).

\bibitem{fan07}
{S.L. Fan and F. Zhong}, Phys. Rev. {\bf 76},  041141  (2007).

\bibitem{lin08}
{S.Z. Lin and B. Zheng}, Phys. Rev. {\bf E78},  011127  (2008).

\bibitem{bag08}
{G. Baglietto and E.V. Albano}, Phys. Rev. {\bf E78},  021125
(2008).

\bibitem{ara07}
{E. Arashiro, J.R. Drugowich de Felicio, and U.H.E. Hansmanno}, J.
Chem. Phys.  {\bf 126},  045107  (2007).

\bibitem{zho08}
{N.J. Zhou and B. Zheng}, Phys. Rev. {\bf E77},  051104  (2008).

\bibitem{he09}
{Y.Y. He, B. Zheng, and N.J. Zhou}, Phys. Rev. {\bf E79},  021107
(2009).

\bibitem{nat01}
{T. Nattermann, V. Pokrovsky, and V.M. Vinokur}, Phys. Rev. Lett.
{\bf 87},  197005  (2001).

\bibitem{kle07a}
{W. Kleemann}, Annu. Rev. Mater. Res. {\bf 37},  415  (2007).

\bibitem{bra94}
A. Bray, Adv. Phys. {\bf {43}},  357  (1994), and references
therein.

\bibitem{bra02}
{A.J. Bray}, Adv. Phys. {\bf 51},  481  (2002).

\bibitem{alb02}
{E.V. Albano and G. Saracco}, Phys. Rev. Lett. {\bf 88},  145701
(2002).

\bibitem{car05}
{S. Caracciolo, A. Gambassi, M. Gubinelli, and A. Pelissetto}, Phys.
Rev. {\bf  E72},  056111  (2005).

\bibitem{smi08}
{T.H.R. Smith, O. Vasilyev, D.B. Abraham, A. Maciolek, and M.
Schmidt}, Phys.  Rev. Lett. {\bf 101},  067203  (2008).

\bibitem{car04}
{S. Caracciolo, A. Gambassi, M. Gubinelli, and A. Pelissetto}, Phys.
Rev. Lett.  {\bf 92},  029601  (2004).

\bibitem{alb04}
{E.V. Albano and G. Saracco}, Phys. Rev. Lett. {\bf 92},  029602
(2004).

\bibitem{ale94}
{F.J. Alexander, D.A. Huse, and S.A. Janowsky}, Phys. Rev. {\bf
B50},  663  (1994).

\bibitem{maj94}
{S.N. Majumandar, D.A. Huse, and B.D Lubachevsky}, Phys. Rev. Lett.
{\bf 73},  182  (1994).

\bibitem{god04}
{C. Godreche, F. Krzakala, and F. Ricci-Tersenghi}, J. Stat. Mech. -
Theory  Exp.  P04007  (2004).

\bibitem{sir04}
{C. Sire}, Phys. Rev. Lett. {\bf 93},  130602  (2004).

\bibitem{zho09}
{N.J. Zhou, B. Zheng, and Y.Y. He}, preprint  (2009).

\end{thebibliography}

\end{document}